\documentclass[11pt,a4 paper]{article} 
\usepackage{amsmath,amscd,amsfonts, graphics, color}
 \usepackage{amssymb, latexsym, framed, rotating} 
\usepackage{setspace}
 \usepackage{authblk}
\makeatletter 
 \renewcommand\@biblabel[1]{#1} 
\makeatother %
\makeatletter
\let\@fnsymbol\@arabic
\makeatother\usepackage{comment}
\usepackage{lipsum}
\usepackage{fullpage}
\usepackage[utf8]{inputenc}
\usepackage{graphicx}
\graphicspath{ {./docu_pictures/} }
\usepackage{amssymb}
\usepackage{forest}
\usepackage{lscape}
\usepackage{fancyvrb}
\usepackage{amsmath, amsthm}
\usepackage{multirow}
\usepackage{rotating}
\usepackage{spverbatim}
\usepackage{tikz}
\usetikzlibrary{trees}
\usepackage[english]{babel}

\usepackage{geometry}
\usepackage{parskip}

\begin{document}
\title{Estimation of survival functions for events based on a continuous outcome: a distributional approach}
\author[1,2,*]{Odile Sauzet}

\affil[1]{Department of Business
Administration and Economics, Bielefeld University, Bielefeld, Germany;}
\affil[2]{Department of Epidemiology \& International Public Health, Bielefeld School of Public Health (BiSPH), Bielefeld University, Germany}

\affil[*]{Corresponding author: Odile Sauzet Universität Bielefeld
Postfach 10 01 31
D-33501 Bielefeld, odile.sauzet@uni-bielefeld.de}
\date{}

\maketitle

\begin{abstract}The limitations resulting from the  dichtomisation of continuous outcomes have been extensively described. But the need to present results based on binary outcomes in particular in health science remains. Alternatives based on the distribution of the continuous outcome have been proposed. Here we explore the possibilities of using a distributional approach in the context of time-to-event analysis when the event is defined by  values of a continuous outcome above or below a threshold. For this we propose in a first step a distributional version of the Kaplan-Meier estimate of the survival function based on repeated truncation of the normal distribution. The method is evaluated with a simulation study and  illustrated with a case study.\end{abstract}

\section{Introduction}
While the statistical literature has long warned of the negative impact of  dichotomisation \cite{fedorov2009consequences}, there are been efforts from applied statisticians to offer methods allowing the estimation of dichotomous parameters based on the underlying distribution.  First Suissa proposed to use the underlying  normal distribution of an outcome to obtain estimates for the binary outcome resulting from dichotomisation \cite{suissa1991binary}.  This approach has been further developed by various authors proposing  a range of methods for  group comparison for proportions  for normal, skew or gamma distributed underlying distributions \cite{peacock2012dichotomising, sauzet2014estimating, sauzet2015, sauzet2018distdichor}, adjusted for covariates \cite{sauzet2016distributional, chen2019using}. The latter have been further extended using a Bayesian approach \cite{jiang2016dichotomizing}, or for count data with zero inflation \cite{preisser2016logistic}. These methods can be part of the assessment of efficacy \cite{kieser2014statistical} or safety in clinical trials  \cite{chis2021improving}

Time-to-event data is per definition based on a binary variable, however  the "event" may be operationalised  as a value of a continuous variable above or under a threshold. For example measures taken on blood are typically continuous but are categorized as normal/abnormal based on some normal range values. The need for an increased use of time-to-event data in safety analysis has been emphasised \cite{cornelius2022improving, phillips2020statistical}. Other area of application include intensive medicine data which contain time-series of continuous blood and other cardiovascular measurements \cite{imhoff2002pattern}. 

Using the underlying continuous distribution of the events to estimate a survival function rather than the observed events  could offer some substantial advantages. Alongside a gain in power due to smaller standard errors, as for other distributional methods, it provides the possibility to obtain estimates also at time-points for which no events are observed. 

In this paper we propose a framework for a distributional survival analysis based on the distribution of (health) outcomes at each time-points for which measures are available. As a first step in this approach, we show how to obtain distributional estimates for the survival function by following a similar approach to the Kaplan-Meier estimates. We present first the theoretical basis of the work. Then we propose to apply the method to normally distributed data and  suggest an approximation of the repeated truncation of the normal distribution by distributions of the skew-normal family \cite{azzalini2013skew}. The method is evaluated with a simulation study and we present a case study using the demo-dataset of the MIMIC-III clinical database \cite{goldberger2000physiobank}. 

\section{Distributional Kaplan-Meier Estimator of survival functions}
\subsection{Estimation of survival function based on the repeat-truncated distribution of the outcome}
In the following we assume that an outcome $Y(t)$, function of the continuous time,   is measured at regular intervals and that the outcome has a known parametric distribution for which a distribution of successive truncation at a fixed threshold (cut-point) is known. We will use the term of repeat-truncation in the sequel. Moreover we assume that an event of interest occurs when the outcome reaches a fixed threshold and that this is always observed (no interval censoring occurs).

a. Distributional assumption on the data: 

Let $(X_i)$ be a times series such that at each time point $t_i\geq 0$, $X_i$ follows a known parametric continuous distribution with parameters $\theta_i$ and such that the distribution of the repeated truncation $Y_i$ satisfies $$P(Y_i<y)=P(X_i<y| X_{i-1}<x_{0},X_{i-2}<x_0,...,X_{0}<x_{0}, \theta_i), \text{ for } i>1 $$ and $$P(Y_1<y)=P(X_1<y| X_{0}<x_{0}, \theta_1) $$

In that case we say that $Y_i$ follows a $i^{th}$ repeat-truncated distribution at cut-point $x_0$.

\begin{figure}[h!]
\includegraphics[width=15cm]{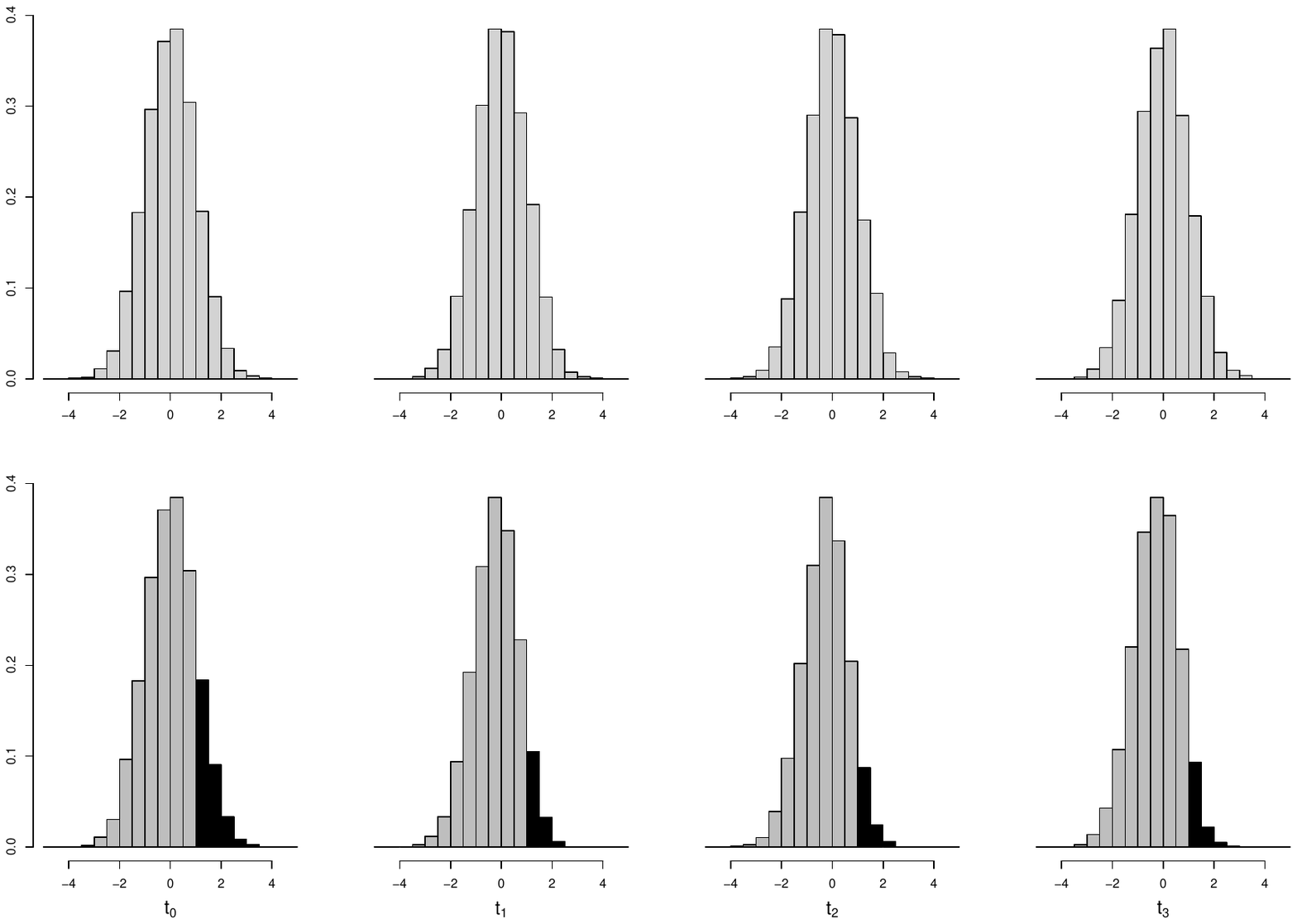}
\caption{Histogram of observed and unobserved values (above, normally distributed) and observed values only (below) due to truncation (cut-point = 1) at previous time-points. The blackened values below indicate the truncated values on the following time-point.  }
\label{trunc}
\end{figure}

The truncation procedure is illustrated in Figure \ref{trunc} for four time points.

\vskip 5pt
b. Definition of population at risk and estimation of distribution parameters for the outcome at time $i$:

 The survival function $S(t)$ is estimated based on the population at risk at time $t$. The population at risk is defined  by all those who have not had an event at any time point preceding $t$ and are being observed at time $t$.

c. Estimation of $S(t)$ at time $t$

The survival function is estimated at only a finite number of time points at which measures are available. Unlike conventional Kaplan-Meier estimation, the survival function can be estimated at time points at which no events are observed
\cite{kaplan1958nonparametric}.
At time $t_i$ the survival function is the product limit estimate  of the survival function at $t_{i-1}$ and of the distributional proportion of the population with an outcome measure above the threshold as estimated above.
$$\hat{S}(t_i)=\hat{S}(t_{i-1})P(Y_i>x_0)$$

An  estimate of  of $P(Y_i>x_0)$ based on the parametric  distribution of $Y_i$ can be obtained if the quantile of the distribution are known and then 

$$\hat{S}(t_i)=\prod_{j=1}^{i}P(Y_j>x_0)$$

In the sequel the estimate for $P(X_i>x_0)$ will be noted $\hat{p}_i$.

An approximate estimation of the standard error of $\hat{p}_i$ can be derived from the distribution of $Y_i$ using the delta method if the distribution of $Y_i$ is known in closed form. Alternatively a bootstrap standard error can be used.

d. Estimation of the standard error of $\hat{S}(t)$

Following the same principles as for the Greenwood formula, we obtain an approximate estimation of the standard error of $\hat{S}(t)$ using twice the delta method which states that 
if $\sqrt{n}|Z_n-\theta|$ converges towards a normal distribution with mean 0 and variance $\sigma^2$, then 
under regularity conditions at $\theta$ for a function $g$, then $\sqrt{n}|g(Z_n)-g(\theta)|$ converges toward $N(0, \sigma^2g'(\theta)^2]$

We pose
$LS(t)=\ln(\hat{S}(t))$.
Following the delta method we have that
$$LS(t_i))=\sum_{j=1}^{i}\ln(1-{\hat{p}}_j)$$
$$\text{Var}(LS(t_i))=\sum_{j=1}^{i}\text{Var}(\hat{p}_j)\ln'(1-{\hat{p}}_j)^2\simeq\sum_{j=1}^{i}\frac{\text{Var}(\hat{p}_j)}{(1-\hat{p}_j)^2}$$ 

Using the delta method a second time to retrieve $S(t)$ we have
$$\hat{S}(t)=\exp((LS(t_i)))$$
so that $$\text{Var}(\hat{S}(t_i))=(\exp((Y(t_i))')^2\text{Var}(LS(t_i))=\hat{S}(t_i)^2\text{Var}(LS(t_i))$$
We therefore obtain a distributional Greenwood formula for the variance of the distributional estimate of the survival function $S(t)$
$$\text{Var}(\hat{S}(t_i))=\hat{S}(t_i)^2\sum_{j=1}^{i}\frac{\text{Var}(\hat{p}_j)}{(1-p_j)^2}$$

\subsection{Approximating the distribution of repeated truncation of a normal distribution}\label{esn}

In this section, we examine the case when the outcome is normally distributed. The aim is to find out if the distribution of a repeat-truncated normal distribution can be approximated by a known parametric distribution. We propose two candidates, the extended skew-normal distribution and the skew normal distribution.

\subsubsection{Extended skew-normal distribution}

A good candidate to approximate the multiple truncation would be is the extended skew-normal distribution (esn) described by Arnold and colleagues  \cite{arnold2002skewed, arnold1993nontruncated} following Azzalini \cite{azzalini2005skew} (see also \cite{azzalini2013skew})
which we briefly describe in our setting:

 Let $X_t\sim N(\mu_{t},\sigma^2)$ and $X_{t-1}\sim N(\mu_{t-1},\sigma^2)$ be two normally distributed random variables with correlation $\delta$ (such that they form a bi-variate normal distribution). Then the conditional probability $$P(X_{t}\leq x | X_{t-1}\leq x_0)$$ is provided by a esn distribution with  the density distribution function :

$$\varphi(x; \mu_t,\sigma^2,\alpha, \tau)=\varphi\left(\frac{x-\mu_t}{\sigma}\right)\frac{\Phi(\tau\sqrt{1+\alpha^2}+\alpha \frac{x-\mu_t}{\sigma})}{\sigma\Phi(\tau)}$$

where $\alpha$ is the skweness parameter and $\tau$ the so-called extension parameter that we will call in our context the threshold parameter (which is used for the truncation).

 This makes this distribution a good candidate for an approximation of the distribution for repeated truncation because it is the exact distribution for the first truncation. However, this distribution necessitates the estimation of four parameters. This is  difficult to achieve in particular for small sample sizes.  Therefore we make another  more practicable suggestion .

\subsubsection{Approximation with a skew-normal distribution}

We propose to  approximate the distribution by another distribution describing truncation of the normal distribution: the skew-normal ($sn$) distribution.

The $sn$ distribution has three parameters and is defined by the following density distribution function: 
$$\varphi(x,\mu_t,\sigma^2, \alpha)=\frac{2}{\sigma}\varphi{\left(\frac{x-\mu_t}{\sigma}\right)}\Phi\left(\alpha\frac{x-\mu_t}{\sigma}\right)$$

where $\alpha$ is the skewness parameter. If $\alpha=0$, the distribution is normal. The adequacy of this distribution to approximate a repeated truncated normal distribution will be assessed in the simulation study. 

The standard error for the estimate of $\hat{p}_i$ can be obtained either using bootstrap or the delta method. The approximate delta method  standard error for $\hat{p}_i$ was derived in \cite{sauzet2015} and is given by: $$se(\hat{p}_i)=\frac{2}{\sqrt{\pi}}\varphi{\left(\frac{x_0-\mu_t}{\sigma}\right)}\Phi\left(\alpha\frac{x_0-\mu_t}{\sigma}\right)$$

\begin{figure}[h! ]
	\includegraphics[width=14.5cm]{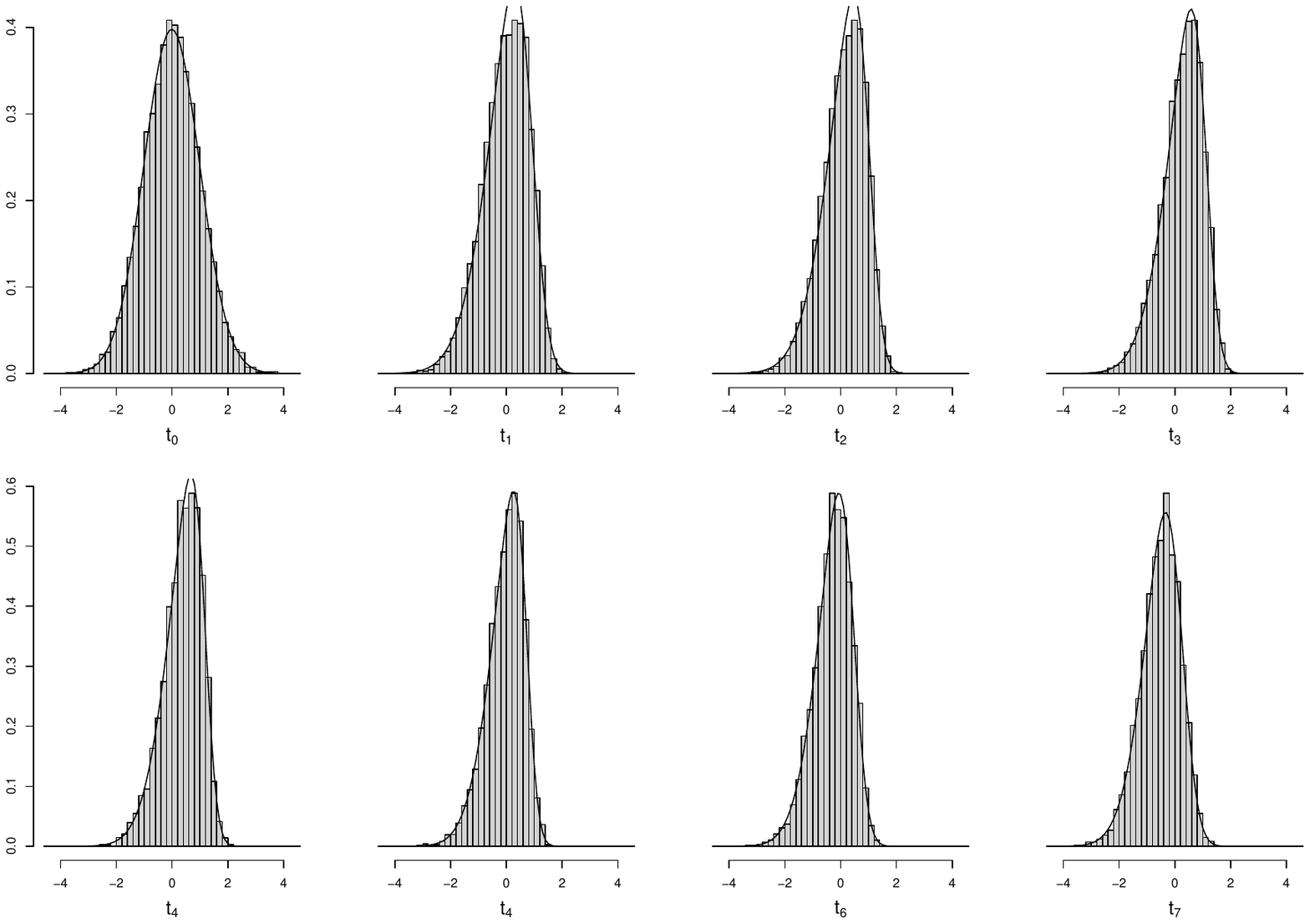}
	
	\caption{Histogram of simulated 7 occurrences of repeat-truncation, cut-point=0, correlation between measures= 0.95  }\label{trunc1}
\end{figure}

\begin{figure}[h! ]
	\includegraphics[width=14.5cm]{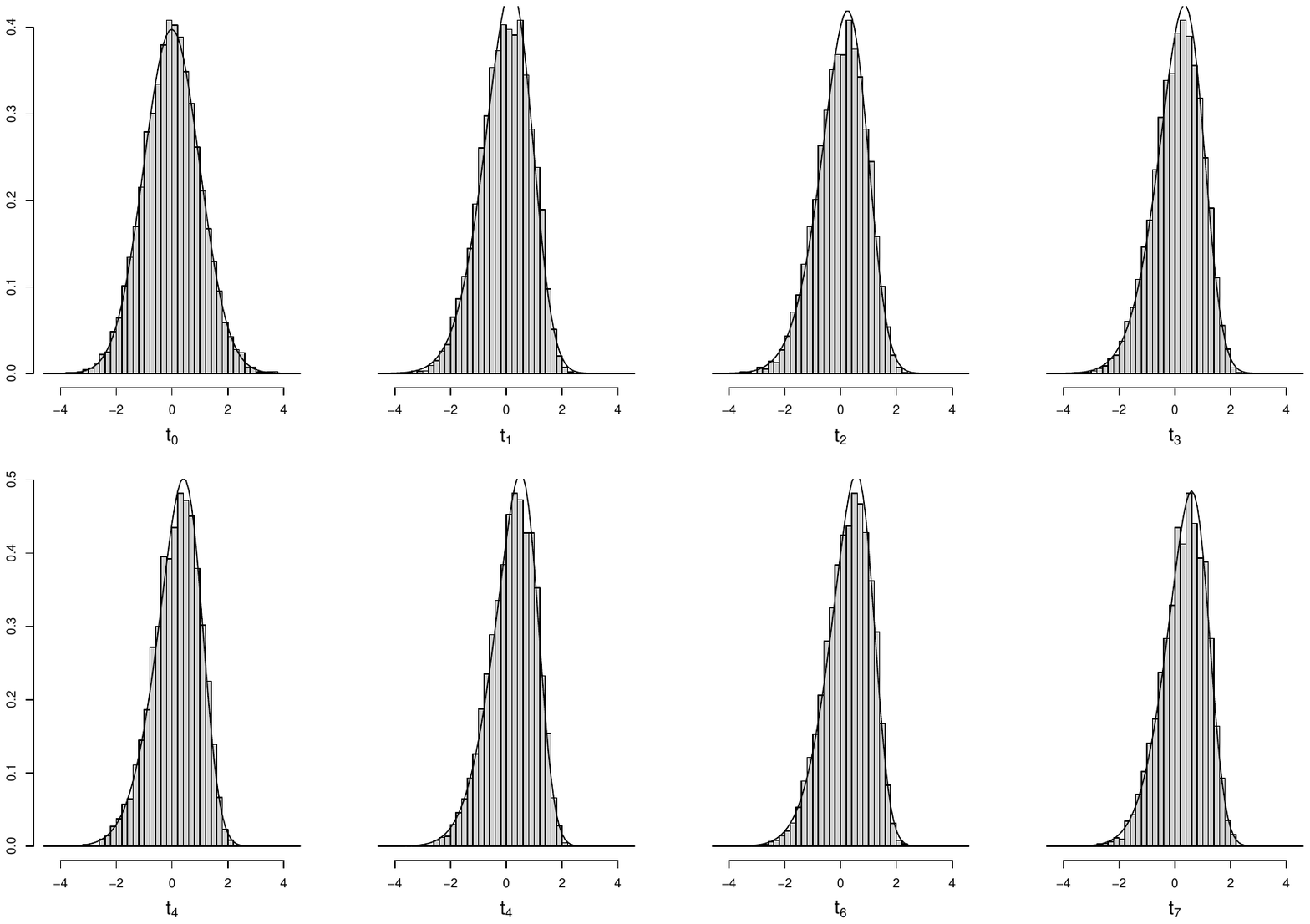}
	
	\caption{Histogram of simulated 7 occurrences of repeat-truncation, cut-point=1.5, correlation between measures= 0.95  }
\label{trunc2}
\end{figure}

We illustrate the fit of the sn-distribution for repeated truncated normal distribution in Fig. \ref{trunc1} (cut-point 0, cor. 0.95) and Fig. \ref{trunc2} (cut-point 1.5, cor. 0.95) in which we compare the distribution function with the histogram of simulated repeat truncated data.


\section{Simulation study}

The aim of the simulation study is to evaluate the reliability of a survival function estimated using skew-normal distribution is a range of conditions. The simulation study consist in the three following sections 

\begin{itemize}
	\item Identification of reference values of the survival function
	\item Simulation scenarios
	\item Evaluation of the results
	
\end{itemize}
\subsection{Simulation scenarios}
We simulated 10 time-points using a multivariate normal distribution with a range of correlation matrices and vectors of means. The data was then truncated inductively such that at every time-point $i$ starting from 1, we replaced by NAs all subsequent observations (time-points $j>i$ ) for participants with values over the cut-point. 
The ranges of values for different scenarios are provided with the vectors of means  in Table \ref{mumatrix}. From the range of simulation scenario, we obtained a wide range of values for S(t) and a wide range of sample size available to estimate them.

\begin{table}[ht]
\centering
\resizebox{12cm}{!} {\begin{tabular}{llllllllllllll}
  \hline\\[-5pt]
 \multicolumn{3}{l}{Scenarios  / Time-points}& 1 & 2 & 3 & 4 & 5 & 6 & 7 & 8 & 9 & 10 \\ 
  \hline\\[-5pt]
\multicolumn{3}{l}{No changes} & 0.00 & 0.00 & 0.00 & 0.00 & 0.00 & 0.00 & 0.00 & 0.00 & 0.00 & 0.00 \\ 
  \multicolumn{3}{l}{Constant increase} & 0.00 & 0.10 & 0.20 & 0.30 & 0.40 & 0.50 & 0.60 & 0.70 & 0.80 & 0.90 \\ 
 
   \hline\\[-5pt]
\multicolumn{3}{l}{Sample sizes}&\multicolumn{4}{l}{Correlations}&&&\multicolumn{4}{l}{Cut-points}\\\\[-8pt]
100&500&1000&0.95& 0.75&0.5&0.20&&&0&0.5&1&1.5\\\hline
\end{tabular}}
\caption {Vectors of means for die 10 time-points and other simulation parameters }\label{mumatrix}
\end{table}
Using the resulting dataset, we obtained the Kaplan-Meier estimate and the $sn$-distributional estimate of the survival function. We also computed the distributional and bootstrap (fixed to 500 repetitions after calibration) standard errors as well as if the reference value (see Section \ref{refval} was covered by the 95\% confidence interval (based on the assumption of normal distribution of the estimates). 

All our simulalion scenarios used without loss of generality a cut-point in the upper tail of the distribution of the outcome. For each scenarios, 500 datasets are generated. The simulation study (and case study below) were perform with $R$ \cite{rcore} using the package $sn$ \cite{rsn}

\subsection{Identification of reference values}\label{refval}

The evaluation of a statistical method should ideally be based on a comparison to true known values. However as discussed in Section \ref{esn} only an approximation of the distribution of the truncated outcome can be obtained, therefore  as alternative we obtained for every simulation scenarios a Kaplan-Meier estimate of the survival function from a very large sample as it is asymptotically unbiased \cite{kaplan1958nonparametric}.  For this, we simulated 500 datasets with a sample size of 80 000.

The reference values are given in Table 1 of the supplementary material. A visible  bias (difference between Kaplan-Meier and {\it sn}-distributional of $>0.005$) is only present when  subsequent measures are almost perfectly correlated (0.95) and the threshold is in the extreme  tail of the distribution ( cut-point = 1 or 1.5). 

For all other scenarios, the large sample approximation with the $sn$ distribution is good enough to estimate the distributional survival function in large sample.

\subsection{Results}

\subsubsection{Comparison Kaplan-Maier - distributional Kaplan-Meier}

We compare the two approaches in the two aspects for which a distributional approach may be beneficial:  comparison of standard errors and number of time points at which estimations are.

\begin{figure}[ht]
	
	\includegraphics[width=10cm]{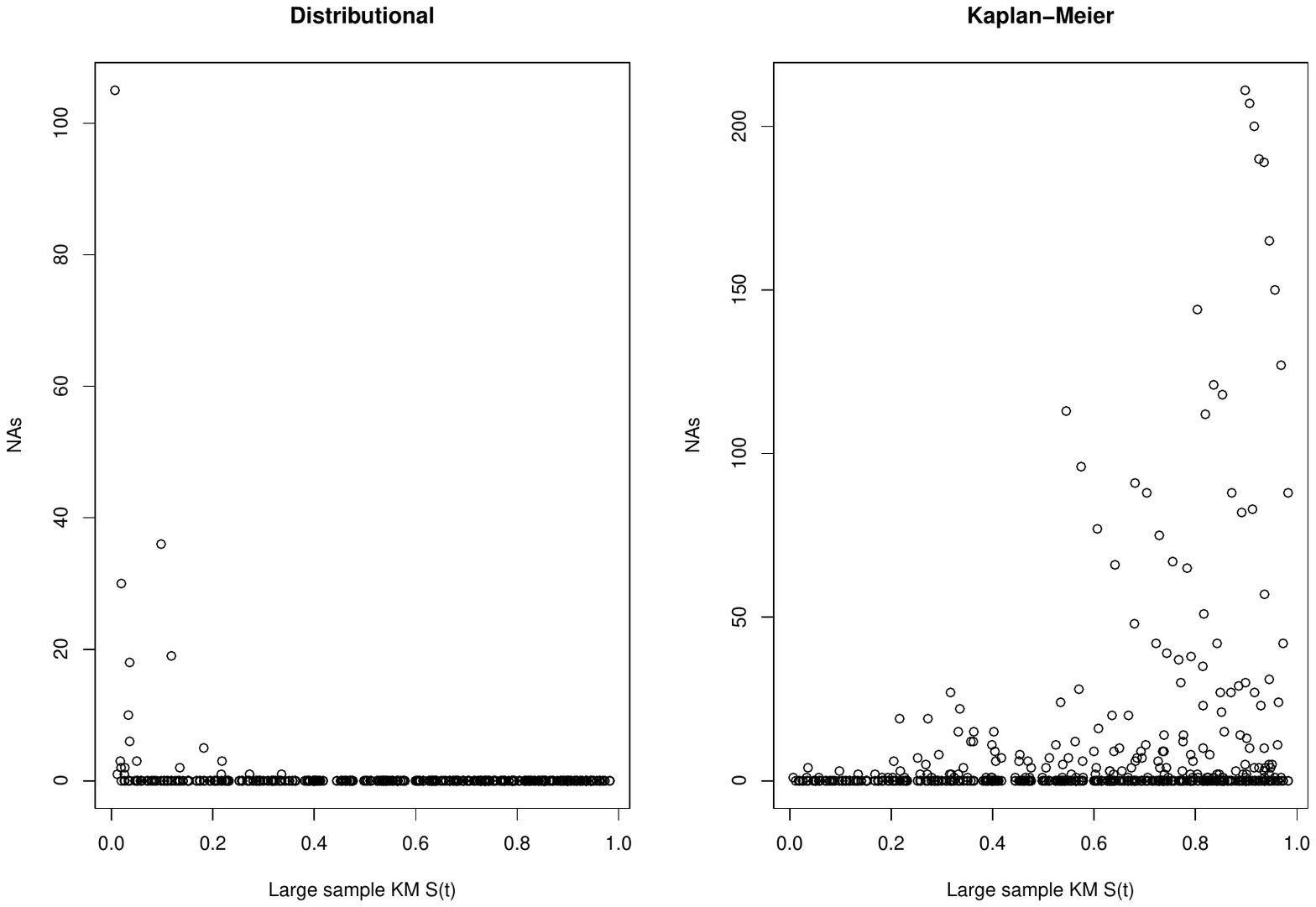}
	
	\caption{Number of missing estimates for S(t) for distributional (left) and Kaplan-Meier (right) approaches }
\label{fig:nast}

\end{figure}

In Figure \ref{fig:nast} the number of missing estimates (from 500) for $S(t)$ relative to the large sample KM-estimate are provided. For all values of $S(t)$ there are missing KM-estimates whereas there are only missing distributional estimates for smaller values of $S(t)$. The latter only occurred if no data was available for estimation (data not shown) when KM-estimates are mostly missing for rare event or if sample size are small. 

\begin{figure}[ht]

	\includegraphics[width=8cm]{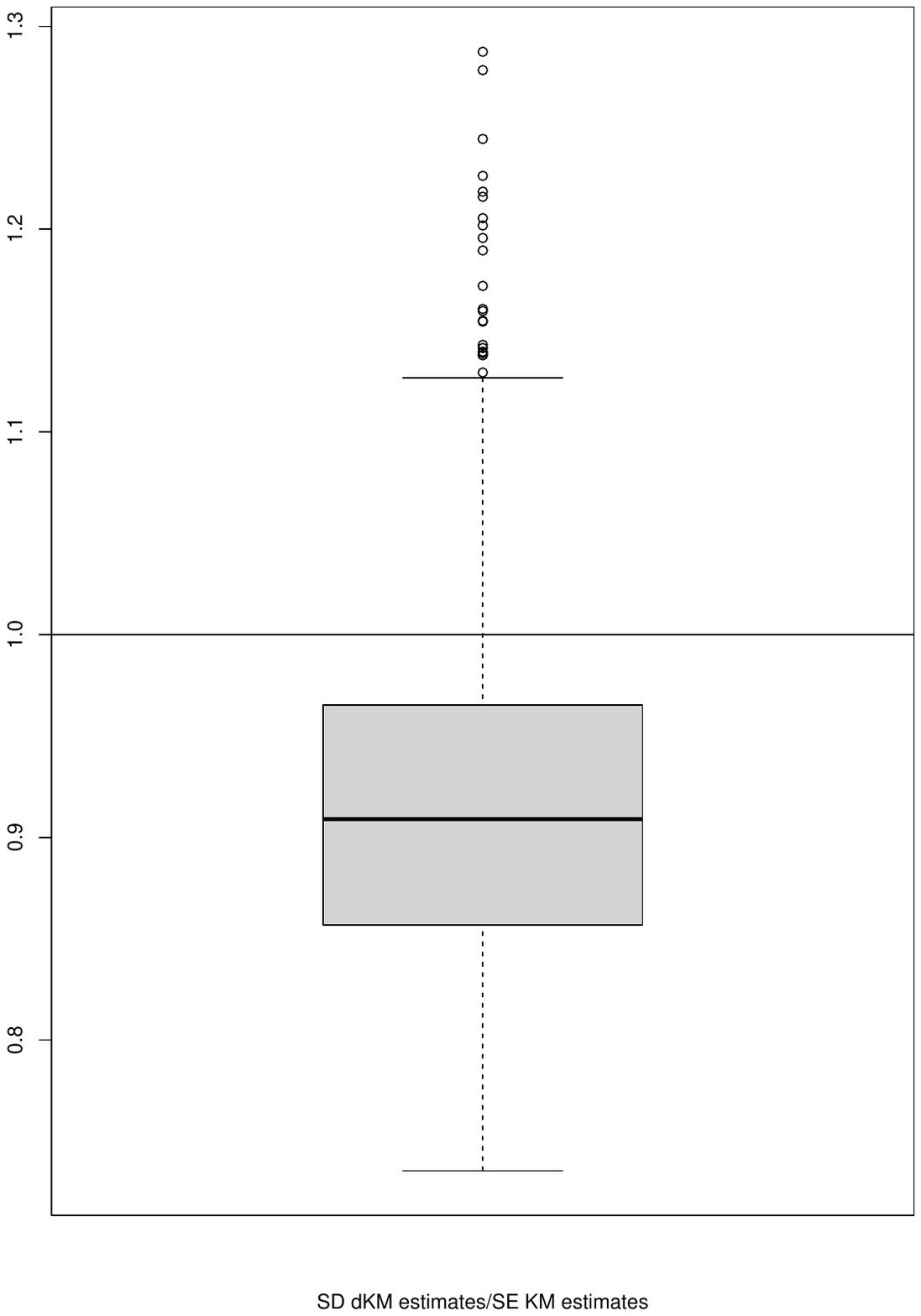}
	
	\caption{Boxplot of the  ratio between the standard deviation of the distributional estimates and the Kaplan-Meier estimates of $S(t)$}
\label{fig:dfse}

\end{figure}

The boxplot for the distribution of ratio of the standard deviation of distributional estimates over the corresponding KM standard errors, Figure \ref{fig:dfse}, shows that overall the distributional estimate have a lower variability than the KM estimates. The extreme values are only obtained for scenarios with very low sample sizes.
\subsubsection{Bias}

\begin{figure}[ht]
	\includegraphics[width=12cm]{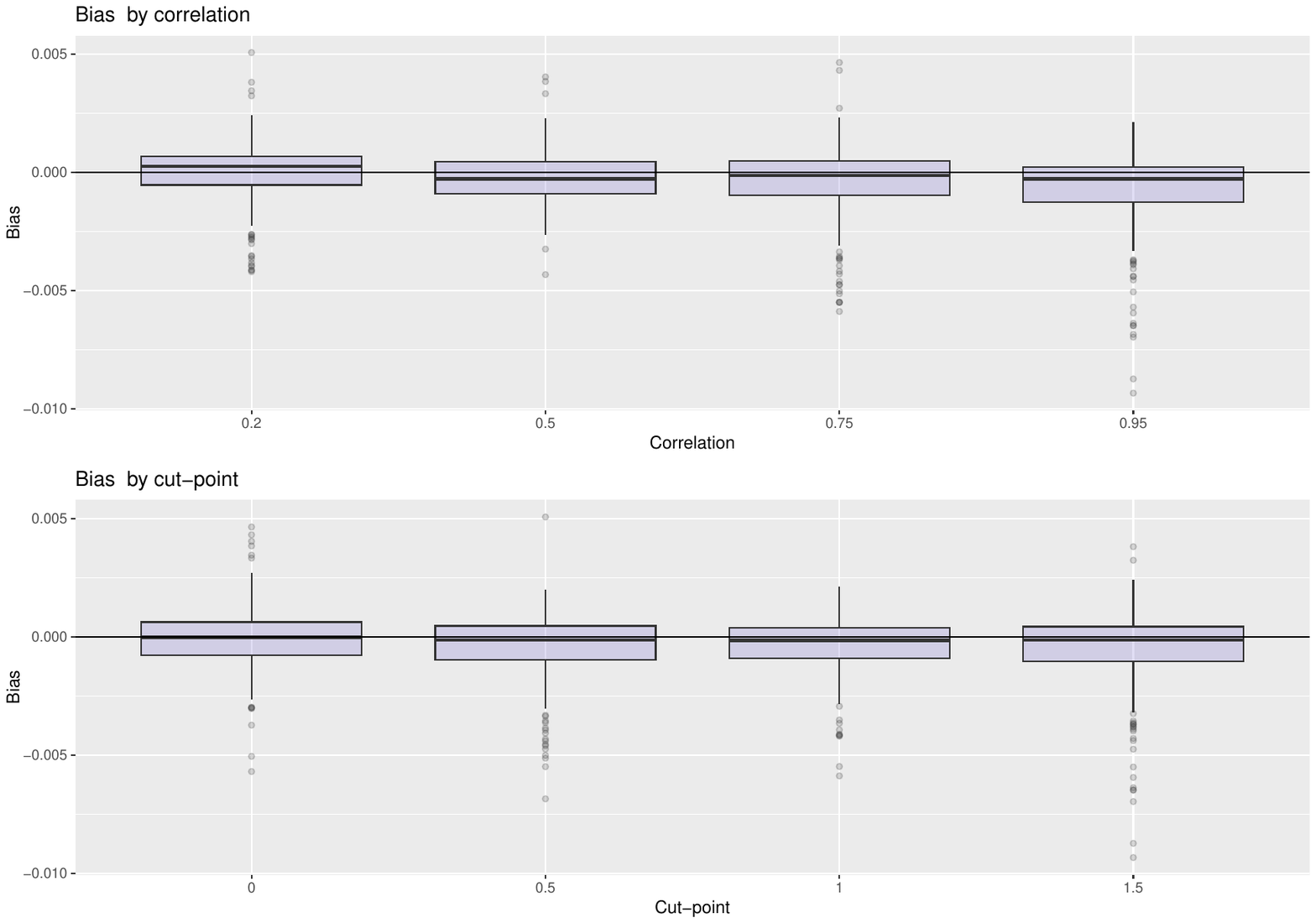}
	
	\caption{Boxplot  of absolute bias of the distribution estimates by correlations and cut-points}
	\label{fig.Nbias}
\end{figure}

\begin{figure}[ht]
	\includegraphics[width=12cm]{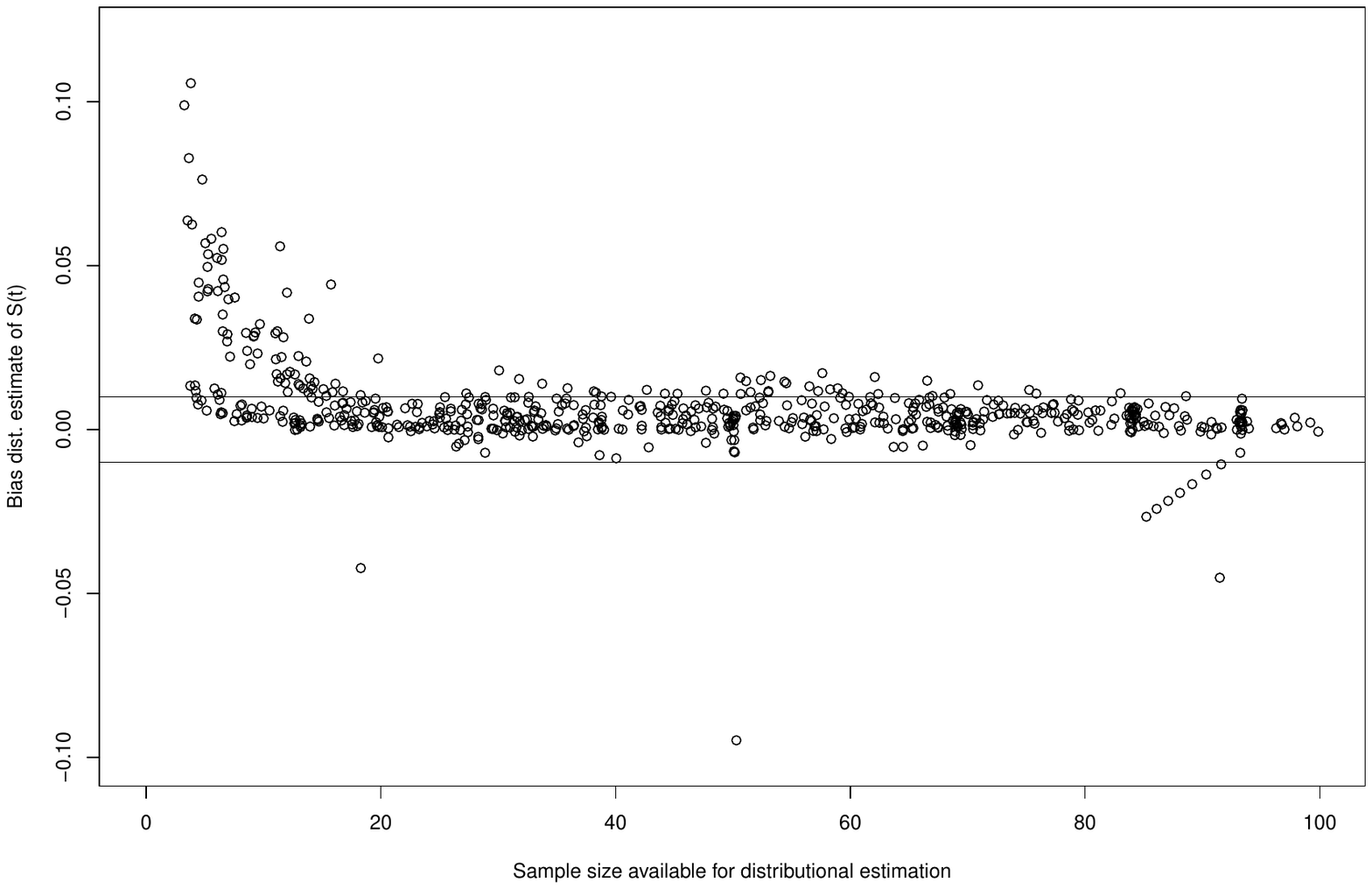}
	
	\caption{Boxplot  of absolute bias of the distribution estimates by correlations and cut-points}
	\label{fig.Nbias100}
\end{figure}

Figure \ref{fig.Nbias} shows a Boxplot  of absolute bias (mean difference between estimated S(t) and large sample KM estimate) of the distribution estimates by correlations and cut-points. These distributions are not directly comparable with each other because difference scenarios lead to different sample size over time, which affects the quality of estimation differently. Not considering come rare extreme cases (high correlation and cut-point in the tail), the estimates seem relatively unbiased.  A scatterplot of bias by sample size  available for estimation of least 100 (Fig. \ref{fig.Nbias100}) shows that a sample size of at least 20 is required to obtain reliable estimates. 

\subsubsection{Standard errors}

\begin{figure}[ht]
	\includegraphics[width=15cm]{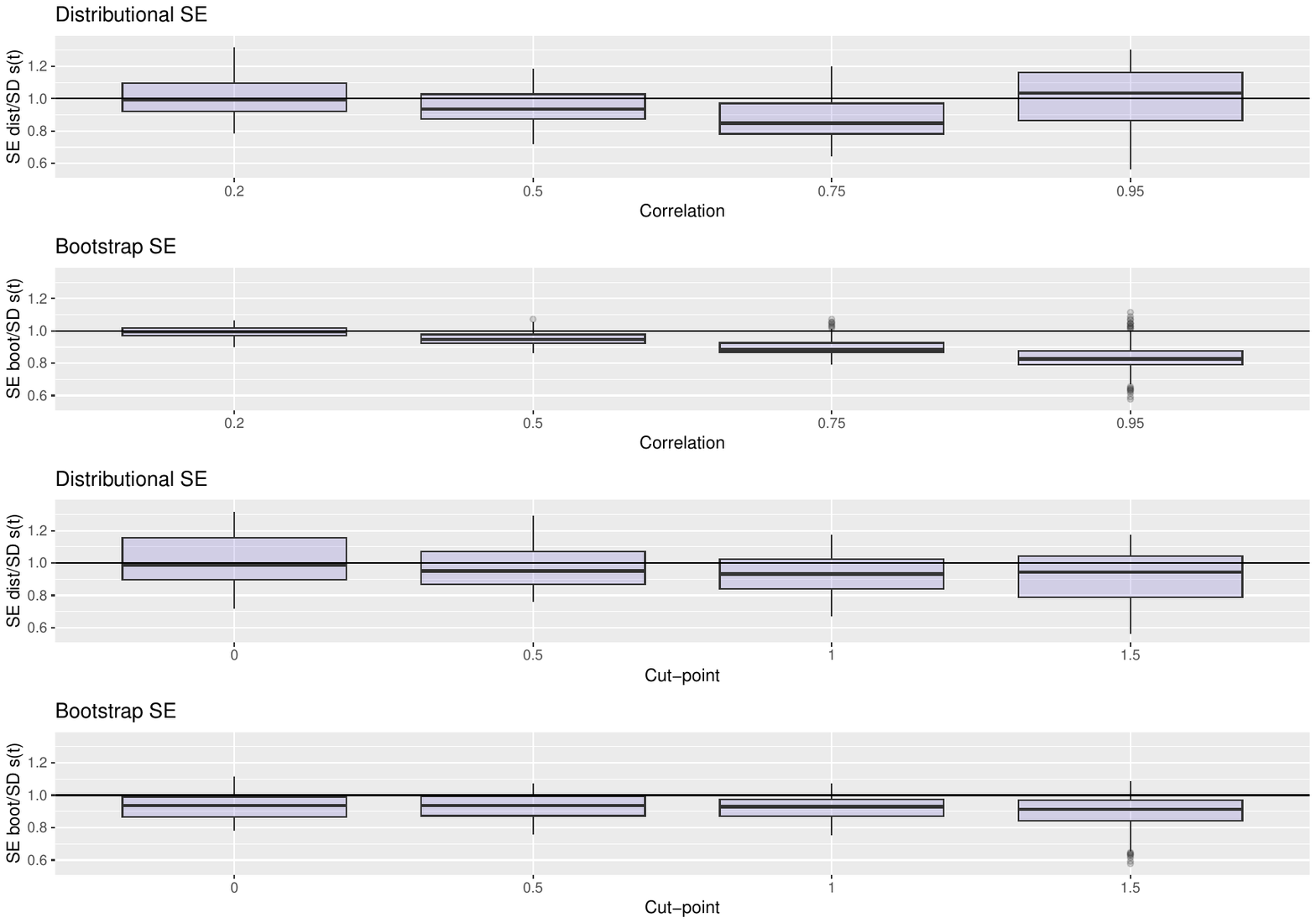}
	
	\caption{Boxplots of the ratio between distributional or bootstrap standard errors (SE) and the standard deviation (SD) of the distributional estimates by correlations and cut-points}
	\label{fig.biasse}
\end{figure}

We focus on the ratio between the estimated standard errors (distributional and bootstrap) and the  standard deviation of the estimates over the 500 simulations. Figure \ref{fig.biasse} represents boxplots of the ratio to the sample size for distributional and bootstrap se. While the ratio for the distributional se seems very close to one there are variations depending of cut-point and correlation between subsequent measures. The bootstrap se is less volatile than the distributional se but seems overall more underestimating the standard deviation of the estimates than the distributional se.

\subsubsection{Coverage of the 95\% confidence interval}
\begin{figure}[ht]
	\includegraphics[width=16cm]{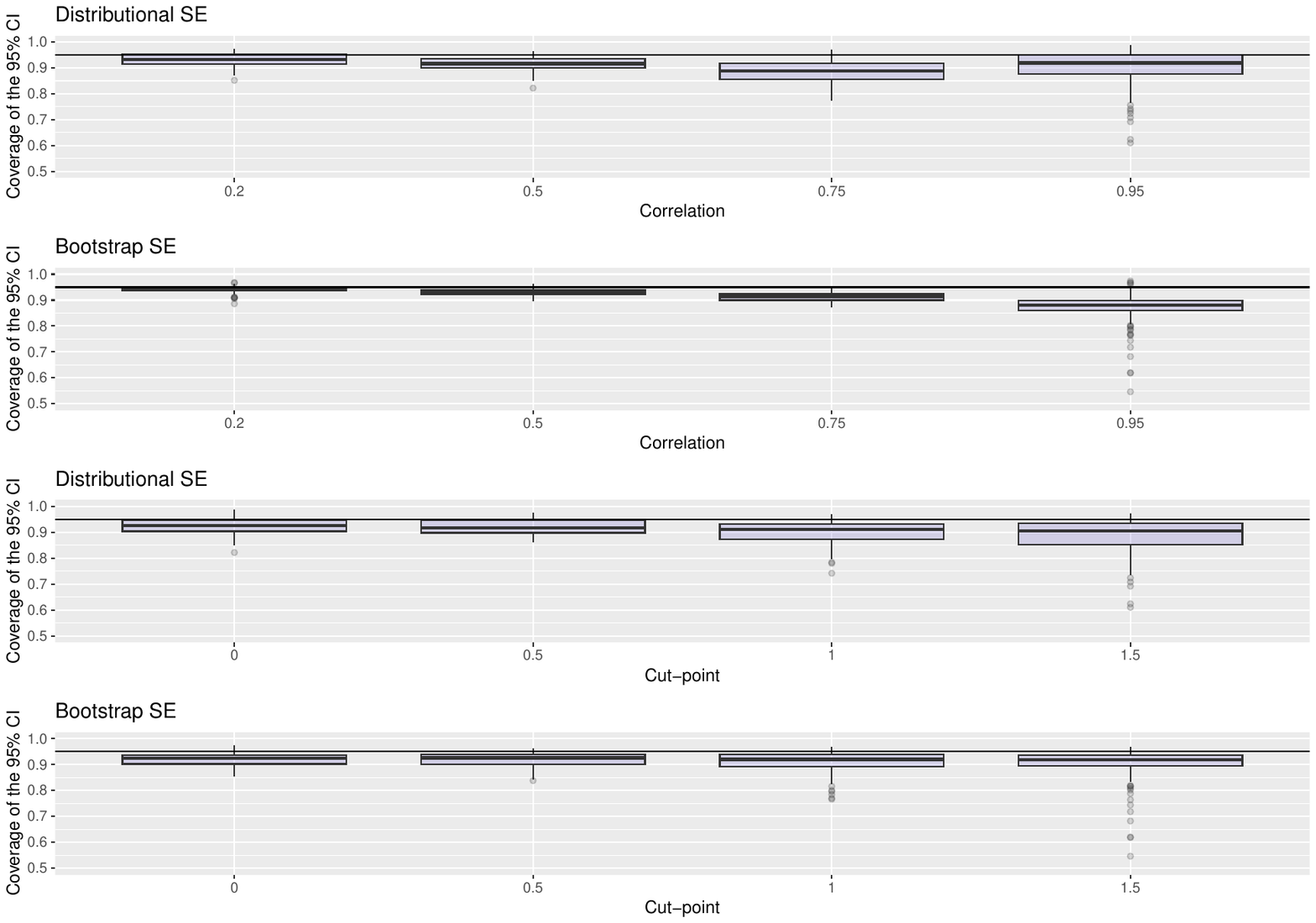}
	
	\caption{Boxplots of the  coverage of the 95\% confidence interval using either distributional or bootstrap standard errors (SE) per correlations and cut-points}
\label{fig.histCovdist}
\end{figure}

The coverage of the 95\% confidence interval was evaluated under the hypothesis of a normal distribution of the estimates of $S(t)$ at each time point at which the survival function was estimated. The coverage was obtained using the distributional standard error (se) and the bootstrap se. In Fig. \ref{fig.histCovdist} we present boxplots of the coverage by correlation and cut-points.  The distribution of coverage reflect the bias and se in which the coverage is mostly lower than 95\% but in most case well above 90\% which makes the method suitable for applications. 

The mean reliability measures over all scenarios by increasing cut-points values and correlation between subsequent measures are summarised in Table \ref{cpcor}. 
\begin{table}[ht]
\centering
\resizebox{12cm}{!} {\begin{tabular}{lrrrrr|lrrrrr}
   \hline&&&&&&&&&&&\\[-5pt]
 CP& MSE & Cov d.& Cov b. & se d. & se b.&Cor & MSE & Cov d.& Cov b. & se d. & se b. \\ 
  \hline&&&&&&&&&&&\\[-5pt]
0 & 0.0014 & 0.932 & 0.921 & 1.117 & 0.969&0.2&0.0007 & 0.937 & 0.944 & 1.083 & 1.017 \\ 
  0.5 &  0.0010 & 0.925 & 0.919 & 1.015 & 0.938&0.5&0.0009 & 0.920 & 0.928 & 0.998 & 0.967 \\ 
  1 & 0.0008 & 0.902 & 0.912 & 0.931 & 0.922 &0.75&0.0011 & 0.888 & 0.912 & 0.890 & 0.904 \\ 
  1.5 & 0.0006 & 0.846 & 0.877 & 0.895 & 0.894 &0.95&0.0011 & 0.862 & 0.847 & 0.983 & 0.838  \\ 
   \hline
\end{tabular}}
\caption{Reliability measures by increasing cut-points values and correlation. CP: cut-point, MSE: mean squared error, Cov d.: coverage of the 95\% confidence interval using the distributional standard error, Cov b.: coverage of the 95\% confidence interval using the bootstrap standard error}
\label{cpcor}
\end{table}
\section{Case study}

The case study we present is based of the freely available demo-dataset from the  MIMIC-III (Medical Information Mart for Intensive Care) clinical database 
which contains clinical data relating to  Intensive Care Unit (ICU) patients \cite{goldberger2000physiobank}. This demo-dataset is limited  to 100 patients, thus  enabling the illustration of the distributional Kaplan-Meier survival estimation on a dataset with small sample size.

We chose the values of bicarbonate concentration in blood which is approximately   normally distributed. The normal values range is given as 22 to 32 mmol/L in \cite{rochencyclo} but other sources provide a lower threshold of 29 \cite{raphael2016bicarbonate}. For illustration purposes, we consider three  different values for the cup-point : 29, 31, 32. 

We operationalised  the data  in the following way: patients have between one and two daily measures registered, we therefore defined the time unit to half days. Each patient is censored at the last half day for which an uninterrupted series of measures is available without missing values. By doing so we stay by the assumption of no interval censoring. This means that measures available subsequently were not used for the estimation of the distribution parameters. At each half day, the parameters of the {\it {\it sn}} distribution were obtained using all the data available at this time point including those with values over the threshold which will be removed for subsequent time points.

Data were available for a total of 98 patients. We limited the analysis to four days (8 time-points). After removing patients with a bicarbonate value above the threshold at baseline, we had 90 patients included in the analysis for a threshold of 32, 88 for a threshold of 31 and 
73 for a threshold of 29. The correlation between consecutive observation was 0.78 between time 2 and 3 and increased to about 0.90 between time 8 and 7. At all time points the sample size available for estimation remains above 20.

The results of the Kaplan-Meier estimations and the distributional survival estimation are provided in Table \ref{cs32}.

\begin{table}[h!]
	\centering
	\begin{tabular}{rrrrrrrrr}

		\hline\\[-5pt]
\multicolumn{5}{l}{Threshold: 29 mmol/L}\\	time & n risk & n event & $\hat{S}(t)$ & se &n est.*  & $\hat{S}_{dist}(t)$  &se$_{boot}$& se$_{dist}$\\ 
		\hline\\
		2 & 73 & 1 & 0.99 & 0.014 & 73 & 0.99 & 0.011 & 0.010 \\ 
		3 & 62 & 2 & 0.95 & 0.026 & 62 & 0.98 & 0.015 & 0.013 \\ 
	 4 & 54 & 3 & 0.90 & 0.038 & 54 & 0.93 & 0.030 & 0.021 \\ 
	5 &--&--&--&-- &  45 & 0.93 & 0.032 & 0.021 \\ 
		6 &--&--&--&--  & 38 & 0.93 & 0.033 & 0.021 \\ 
	7 & 37 & 2 & 0.85 & 0.049 & 36 & 0.90 & 0.038 & 0.025 \\ 
	8 &--&--&--&-- &  34 & 0.87 & 0.040 & 0.030 \\ 
	\end{tabular}

	\centering
	\begin{tabular}{rrrrrrrrr}
		\hline\\[-5pt]
\multicolumn{5}{l}{Threshold: 31 mmol/L}\\
		time & n risk & n event & $\hat{S}(t)$ & se &n est.*  & $\hat{S}_{dist}(t)$  & se$_{boot}$& se$_{dist}$\\ 
		\hline\\
		
		2 &--&--&--&-- & 88 & 0.99 & 0.008 & 0.007 \\ 
		
		3 & 71 & 2 & 0.97 & 0.020 &71 & 0.98 & 0.010 & 0.008 \\ 
		
		4 & 63 & 2 & 0.94 & 0.029&  63 & 0.96 & 0.017 & 0.013 \\ 
		5 & 54 & 1 & 0.92 & 0.033& 54 & 0.92 & 0.028 & 0.018 \\ 
		
		6 & 46 & 1 & 0.90 & 0.038 &46 & 0.91 & 0.029 & 0.019 \\ 
		7 &--&--&--&--& 42 & 0.90 & 0.030 & 0.021\\ 
		8&--&--&--&--& 42 & 0.89 & 0.030 & 0.023  \\
	\end{tabular}
	\centering
\begin{tabular}{rrrrrrrrr}
	\hline\\[-5pt]
\multicolumn{5}{l}{Threshold: 32 mmol/L}\\
	time & n risk & n event & $\hat{S}(t)$ & se &n est.*  & $\hat{S}_{dist}(t)$  & se$_{boot}$& se$_{dist}$\\ 
	\hline\\
		2 &--&--&--&--  & 90 & 1.00 & 0.005 & 0.003 \\ 
		3 &--&--&--&--  & 72 & 0.99 & 0.007 & 0.003 \\ 
		4 & 65 & 3 & 0.95 & 0.026 & 65 & 0.96 & 0.018 & 0.012 \\ 
	 5 & 55 & 1 & 0.94 & 0.031& 55 & 0.94 & 0.023 & 0.015 \\ 
	6 &--&--&--&--  & 47 & 0.93 & 0.024 & 0.016 \\ 
	7 &--&--&--&--  & 44 & 0.92 & 0.026 & 0.036 \\ 
	8 &--&--&--&--  & 44 & 0.91 & 0.027 & 0.045 \\ 
		\hline
	\end{tabular}
		\caption{Kaplan-Meier and distributional estimates of the survival function for bicarbonate blood concentration of patient in intensive station for a range of threshold. se: Standard error ($dist$: distributional, $boot$: bootstrap) *Sample size available for the estimation of the {\it sn} distribution parameters.}\label{cs32}
\end{table}

Estimation of the survival function: The Kaplan-Meier estimator was obtained for a maximum of four time-points (only two for a threshold of 32), whereas the distributional estimator was obtained for all the 7 post baseline time-points. There are some differences between the two estimators but none of which are not within their respective confidence intervals.

Comparison between standard error based on the distribution parameters and the bootstrap: The bootstrap standard errors (se$_{boot}$) is systematically higher than the distributional standard error se$_{dist}$ in line with what would be expected from the simulation study. Only for the threshold of 32, does the se$_{boot}$ become higher (jumping from 0.016 at time-point 6 to 0.36 at time point 7). This difference  may be explained by increased difficulty to  estimate the quantile in the tail of the distribution.

The se$_{dist}$ were systematically lower than the Kaplan-Meier standard errors. This is also to be expected from the simulation study for which most simulation scenarios provided distributional estimates with less variability than the KM estimates. Distributional se can be slightly underestimated but event even so the distributional estimates of the survival functions are more precise than the KM estimates.

\section{Discussion}

We presented an approach to survival analysis in which we made use of the  distribution underlying  the binary outcome defining the event of interest. We illustrated this approach by proposing a distributional Kaplan-Meier estimation of the survival function. The difficulty of implementing such an approach lies in the determination of the distribution of repeat-truncation of (normal) distribution. In this work we used the skew-normal distribution to approximate repeat-truncation of the normal distribution. Following an evaluation, we saw that overall the method provided reliable estimates. The conclusion of our work include the following:

\begin{itemize}
\item The {\it sn} distribution is a suitable approximation for repeat-truncation of the normal distribution to obtain estimates of $P(t>T|t\geq T)$.
	\item The approach  provides estimates of the survival function at every observed time points whether an even is observed or not.
	\item The distributional standard errors are in most cases smaller than  those derived form the binary Greenwood formula.

\end{itemize}
 
 Through the simulation study we could deduce the following limitations:
  
\begin{itemize}
\item Necessitate al least 20 observations to estimate S(t).

\item For cut-points in the extreme tail and high correlation between the subsequent measures, the {\it sn} distribution does not approximate the  of repeat-truncation well enough to obtain reliable estimates of the survival function.

\end{itemize}

A distributional survival function estimate using the {\it sn} distribution was obtained and shown to provide reliable estimate in application. This approach has many features common to the Kaplan-Meier estimate, in particular, it is based on a product-limit estimator with the only difference that the probability to survive after time $t$ conditional to have survived until time $t$ is a quantile of a repeat-truncated (normal) distribution. In that respect it remains a non-parametric approach as no distributional assumptions are made on the time itself.

In this first simple version of a distributional estimation of the survival function, we made only use of the data at time $t$ to estimate the distribution parameters at time $t$. A direction for further development would be to make use of all the data available at times $<t$ to estimate the distribution parameters in order to reduce the standard errors further. Another direction of research is be to find other distribution to model repeated truncation with parameters estimable with limited data as well as include distributional information on the generation of the event in other survival analysis methods.


\end{document}